\documentclass[aps,pra,twocolumn,amsmath,superscriptaddress,tightenlines,longbibliography]{revtex4}
\usepackage{amssymb}
\usepackage{amsmath}
\usepackage{dcolumn}
\usepackage{graphicx}
\usepackage{mathrsfs}
\usepackage{color}

\usepackage{url}
\usepackage[colorlinks]{hyperref}
\hypersetup{%
    plainpages=true,
    breaklinks=true,
    hypertexnames=false,
    pageanchor=true,
    colorlinks=true,
    linkcolor={blue},
    citecolor={blue},
    urlcolor={blue},
    anchorcolor={black}
}

\begin{document}

\title{Quantum memory and gates using a $\Lambda$-type quantum emitter
coupled to a chiral waveguide}

\author{Tao Li}
\affiliation{Theoretical Quantum Physics Laboratory, RIKEN Cluster for Pioneering Research, Wako-shi, Saitama 351-0198, Japan
 }
\author{{Adam Miranowicz}\footnote{miran@amu.edu.pl}}
\affiliation{Theoretical Quantum Physics Laboratory, RIKEN Cluster for Pioneering Research, Wako-shi, Saitama 351-0198, Japan
 }
\affiliation{Faculty of Physics,  Adam Mickiewicz University,
61-614 Pozna\'{n},  Poland}
\author{Xuedong Hu}
\affiliation{Theoretical Quantum Physics Laboratory, RIKEN Cluster for Pioneering Research, Wako-shi, Saitama 351-0198, Japan
 }
\affiliation{Department of Physics, University at Buffalo, SUNY,
Buffalo, New York 14260-1500, USA}
\author{{Keyu Xia}\footnote{keyu.xia@nju.edu.cn}}
\affiliation{Theoretical Quantum Physics Laboratory, RIKEN Cluster for Pioneering Research, Wako-shi, Saitama 351-0198, Japan
 }
\affiliation{College of Engineering and Applied Sciences, Nanjing
University, Nanjing 210008, China}
\author{{Franco Nori}\footnote{fnori@riken.jp}}
\affiliation{Theoretical Quantum Physics Laboratory, RIKEN Cluster for Pioneering Research, Wako-shi, Saitama 351-0198, Japan
 }
\affiliation{Physics Department, The University of Michigan, Ann
Arbor, Michigan 48109-1040, USA}

\date{\today }

\begin{abstract}
By coupling a $\Lambda$-type quantum emitter to a chiral waveguide,
in which the polarization of a photon is locked to its propagation
direction, we propose a controllable photon-emitter interface for
quantum networks. We show that this chiral system enables the SWAP
gate and a hybrid-entangling gate between the emitter and a flying
single photon.  It also allows deterministic storage and retrieval
of single-photon states with high fidelities and efficiencies.  In
short, this chirally coupled emitter-photon interface can be a
critical building block toward a large-scale quantum network.
\end{abstract}

\maketitle

\section{Introduction}

A quantum network could provide   secure information
distribution protected by the quantum no-cloning
theorem~\cite{kimble2008quantum, zheng2010arbitrary,
PhysRevA.69.052319}. Photons at optical frequencies, which usually
interact weakly with their environment, are natural information
carriers (a quantum bus) in a quantum network, connecting remote
quantum computer nodes~\cite{cirac1997quantum,
ritter2012elementary, pan2012multiphoton, Deng201746} via quantum
emitters. Within this paradigm of a quantum network, a highly
efficient and reliable interface between flying photons and
stationary emitters is a prerequisite for building quantum
networks~\cite{li2013entanglement,hao2015quantum}.

A cavity strongly coupled to a quantum emitter is a well-known
photon-emitter interface. It can accomplish a variety of
elementary quantum information processing (QIP)
tasks~\cite{lodahl2015interfacing, reiserer2015cavity,
RevModPhys.85.623}, ranging from single-photon
sources~\cite{kuhn2002deterministic, keller2004continuous,
holleczek2016quantum, barik2018topological} to quantum
gates~\cite{duan2004scalable, koshino2010, reiserer2014quantum,
tiecke2014nanophotonic, PhysRevA.96.012315, hacker2016photon,
sun2016quantum}, quantum memories~\cite{maitre1997quantum}, and
quantum routers~\cite{PhysRevLett.101.100501,
PhysRevLett.111.103604, PhysRevA.89.013805, PhysRevA.84.033854, hu2016spin,
cao2017implementation,xia2014reversible,
shomroni2014all}.  However, constructing a multinode
quantum network requires an array of strongly coupled cavities in
a cascaded arrangement.  Despite all this remarkable experimental
progress, it remains a difficult technical challenge to connect
different cavities while maintaining the required strong
coupling~\cite{PhysRevLett.102.083601}.


Nanoscale optical waveguides   provide an intriguing alternative to cavities as an efficient interface between a single photon and a single atom.  In a one-dimensional (1D) nanoscale optical waveguide, photon fields are tightly confined in the transverse direction, so that a photon can interact strongly with a nearby atom~\cite{gu2017microwave,shen2005coherent}.  Conversely, atoms can act as quantum scatterers for photons and control the photon propagation along the waveguide.  Near-unity coupling efficiency of a quantum emitter to a photonic-crystal waveguide has been achieved experimentally~\cite{Arcari14}. Moreover, waveguides are naturally scalable and can be easily integrated on a chip to scale up the number of nodes of a quantum network. With these attractive characteristics, coupled waveguide-emitter systems have generated tremendous interest for QIP and quantum
network applications~\cite{lodahl2017chiral, guimond2016chiral, PhysRevA.91.042116, PhysRevA.83.021803, bliokh2014extraordinary,bliokh2015quantum, bliokh2015spin, bliokh2015transverse, young2015polarization, sollner2015deterministic,songPRA2017}.

Recent studies of the coupled atom-waveguide system have revealed
it as a versatile tool for quantum
networks~\cite{lodahl2017chiral, guimond2016chiral,
PhysRevA.91.042116}.  For instance, in a 1D waveguide a two-level
atom strongly coupled to the optical field can be treated as a
mirror with a tunable reflectivity~\cite{PhysRevA.78.063827,
PhysRevA.83.013825, chang2012cavity,
PhysRevLett.113.243601, PhysRevLett.117.133603, chang2007single}.
A multilevel atom allows an even more flexible control of photon
propagation~\cite{witthaut2010photon, PhysRevA.94.053857,
guimond2016chiral, PhysRevLett.111.090502, li2012robust,
PhysRevA.85.043832, PhysRevA.93.013849, harris2016large}. For
example, a V-type atom can act as a single-photon
diode~\cite{PhysRevX.5.041036, xia2014reversible}. A
$\Lambda$-type emitter, resonantly coupled to a single photon
propagating in a 1D waveguide through one of two transitions, can
flip its two ground states and simultaneously  emit a photon
coupled to the other transition with a probability up to
$50\%$~\cite{tsoi2009single}. Even more interestingly, when the
two transitions of a $\Lambda$-type atom are driven by the same
waveguide mode~\cite{witthaut2010photon}, an effective flip of the
two atomic ground states can occur. This process has been
exploited to perform a controlled-phase-flip gate on two
atoms~\cite{ciccarello2012quasideterministic}. By introducing a
Sagnac interferometer and placing the atom at a balanced point, a
single-photon interference via a $\Lambda$-type atom can enable a
deterministic frequency conversion of single
photons~\cite{ bradford2012single}.

In this paper, we propose a photon-emitter interface by coupling a
$\Lambda$-type quantum emitter to a 1D chiral
waveguide~\cite{lodahl2017chiral,
guimond2016chiral,PhysRevA.91.042116}.  The photonic spin and
momentum are locked for photons propagating in a 1D chiral
waveguide~\cite{PhysRevA.83.021803,bliokh2014extraordinary, bliokh2015quantum,
bliokh2015spin, bliokh2015transverse}.  Consequently,   the emitter
transition between each of the ground states and the excited state
selectively couples to the forward- or backward-propagating
photons.  A single photon propagating in such a waveguide thus has
a chiral interaction with the $\Lambda$-type emitter, with
coupling strength dependent on its traveling direction.  Here we
show that such a chiral photon-emitter system, which we refer to
as a multifunctional quantum interface (MQI), can perform a wide
variety of QIP tasks useful for a quantum network.  These tasks
include the SWAP gate and its square root $\sqrt{\textrm{SWAP}}$
(which is an entangling gate) between the emitter and a single
photon, as well as using the emitter as a quantum memory of a
single photon in an arbitrary superposition state.  A key
advantage of this chiral system is that all these quantum
processes could be realized in the exact same setup using a chiral
protocol.  A particular function of the MQI can be selectively
enabled by applying a proper detuning between the emitter and the
input photon.  We show that such a selection is robust against
deviations from ideal scattering conditions, such as a finite
bandwidth of   the photon and a finite effective interaction between
the emitter and the waveguide.
 {A specific quantum function (such as the SWAP gate) can be
triggered if and only if photons are injected into the setup in a
specific direction, while photons injected in the opposite
direction do not cause any effect. This is significantly different
from standard realizations of quantum gates, such as the protocol
for a photon SWAP gate discussed in Ref.~\cite{koshino2010}.}
In short, such a chiral quantum system
could be a truly compact, versatile, and powerful addition to the
development of a complex quantum network.

The rest of the article is organized as follows: In Sec.~\ref{sec2}, we present the single-photon chiral scattering process in a coupled emitter-waveguide system. In Sec.~\ref{sec3}, we propose an MQI based on   single-photon chiral scattering, which enables state swapping, entanglement generation, and quantum memory. In Sec.~\ref{sec4}, we analyze the performance of the MQI for different QIP tasks. Finally, we conclude with a brief discussion and summary in Sec.~\ref{sec5}

\section{Photon scattering by a chiral emitter-waveguide system} \label{sec2}

\begin{figure}
\includegraphics[width=8.5 cm,angle=0]{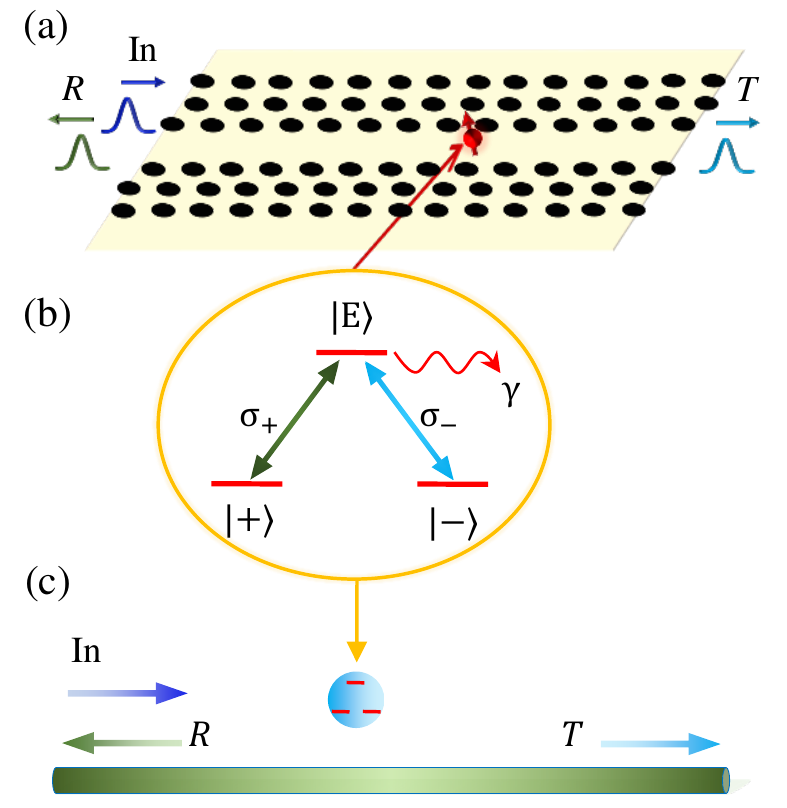}
\caption{(a) Chiral scattering of a single photon by a
$\Lambda$-type emitter coupled to a photonic-crystal waveguide,
(b) energy levels and the circularly polarized ($\sigma_{\pm}$)
dipole transitions of a $\Lambda$-type emitter (decaying with a
damping rate $\gamma$), and (c) chiral scattering of a single
photon by a $\Lambda$-type emitter coupled to the waveguide. Here
$R$ and $T$ represent the reflection and transmission amplitudes
of the scattered output modes. For simplicity, we show only a
photon entering the waveguide from its left-hand side (say, a
forward-propagating photon), although we also consider a
backward-propagating photon. The atom is placed at a point of the
local in-plane (circular) polarization of the electric field in
the waveguide (see Ref.~\cite{lodahl2017chiral} for a pedagogical
description of this configuration). The circularly-polarized
$\sigma_{+}$ ($\sigma_{-}$) dipole transition couples \emph{only}
to the forward- or backward-propagating modes with the coupling
constants $\Gamma_{f}(k)=V_f^2(k)$ or $\Gamma_{b}(k)=V_b^2(k)$, respectively.}
\label{fig1}
\end{figure}

The schematic of our MQI via a chiral scattering process is
depicted in Fig.~\ref{fig1}. Specifically, we consider a single
(natural or artificial) three-level $\Lambda$-type
atom~\cite{buluta2011natural},   e.g., nitrogen-vacancy (NV)
centers in diamond, which is embedded in a photonic crystal
waveguide~\cite{young2015polarization, sollner2015deterministic}
or a nanobeam waveguide~\cite{le2015nanophotonic}. The atomic
transition  $\vert+\rangle\leftrightarrow\vert E\rangle$
($\vert-\rangle\leftrightarrow\vert E\rangle$) is optically driven
by the $\sigma^{+}$ ($\sigma^{-}$) polarized photon.  The chiral
interaction between the atom and the waveguide originates from the
combination of the following properties.  { Through our
investigation below, we refer to the right-moving (left-moving)
waveguide modes as the forward-propagating (backward-propagating)
modes.}   The forward-propagating electromagnetic field with the
wave vector $\overrightarrow{k}$ and the backward-propagating modes
with the  opposite wave vector $(-\overrightarrow{k})$ are related to
each other   via $\overrightarrow{E}_{\overrightarrow{k}}
=\overrightarrow{E}_{-\overrightarrow{k}}^{*}$ due to the
time-reversal symmetry~\cite{lodahl2017chiral,bliokh2014extraordinary,
bliokh2015quantum,bliokh2015spin,bliokh2015transverse}. When
placing the atom in the waveguide at a point  {where the local
in-plane electric field is circularly polarized,} the two
circularly polarized dipole transitions of the atom are coupled to
oppositely propagating modes. The transition, driven by
  a $\sigma_{+}$ ($\sigma_{-}$) polarization, can only be coupled to
the forward-propagating (backward-propagating) $k$-mode with the coupling
constant $\Gamma_{f}(k)=V_f^2(k)$ [$\Gamma_{b}(k)=V_b^2(k)$].

The Hamiltonian $\hat{H}$ is, thus, given by
\begin{eqnarray}\label{eqH}
\hat{H}=\hat{H}_0+\hat{H}_1,
\end{eqnarray}
where the free Hamiltonian $\hat{H}_{0}$ and the interaction
Hamiltonian $\hat{H}_{1}$ can be given in the dipole and
rotating-wave approximation, respectively, as
\begin{eqnarray}
\hat{H}_{0}&=&\Omega_{{_E}}\hat{\sigma}_{_{EE}}
+\sum_{j=\pm}\left[\int_{0}^{\infty}{\rm
d}k\,\omega_{jf}(k)\hat{a}_{jf}^{\dagger}(k)\hat{a}_{jf}(k)\right.\nonumber\\\label{Hamliltonian0}
 &&\left.+\int_{-\infty}^{0}{\rm d}k\,\omega_{jb}(k)\hat{a}_{jb}^{\dagger}(k)\hat{a}_{jb}(k)\right],\\\nonumber
\hat{H}_{1}&=&\!\!\int\!\! {\rm
d}k\left[V_{f}(k)\hat{a}_{+f}^{\dagger}(k)\hat{\sigma}_{+{_E}}
+V_{b}(k)\hat{a}_{-b}^{\dagger}(x)\hat{\sigma}_{-{_E}}+{\rm h.c}\right],\\
\label{Hamliltonian1}
\end{eqnarray}
where, for simplicity, we set the group velocity $c=1$ and
$\hbar=1$, the frequency of a single photon is assumed to be
identical to its absolute value of the   wave vector
($\omega_k=|k|$), and the energy of the atomic ground states
$|\pm\rangle$ is assumed to be zero. Moreover,
$\hat{\sigma}_{iE}=\vert i\rangle\langle E\vert$ with $i=\pm,E$
are the atomic transition operators, and $\Omega_{E}$ is the
optical-transition frequency of the atom. This chiral Hamiltonian,
given in Eq.~(\ref{eqH}), is a generalization of standard
Hamiltonians describing a nonchiral interaction of a multimode
electromagnetic field and a two-level~\cite{carmichaelQO} or
three-level~\cite{tsoi2009single} atom. The creation operator
$\hat{a}_{jf}^{\dagger}(\omega)$
[$\hat{a}_{jb}^{\dagger}(\omega)$] creates a $j$-polarized photon
of frequency $\text{\ensuremath{\omega}}$ in the forward-propagating
(backward-propagating) modes along the waveguide. As considered in
previous works~\cite{gu2017microwave,shen2005coherent,Arcari14,
lodahl2017chiral,guimond2016chiral,PhysRevA.91.042116}, a
linearized dispersion relation of photons around the frequency
$\omega_{0}=|\pm \overrightarrow{k}_{0}|$ of an input photon is
used here with
$\omega_{j\!f}(\overrightarrow{k})=\omega_{0}+k-k_{0}$ and
$\omega_{jb}(-\overrightarrow{k})=\omega_{0}-k+k_{0}$. In
addition, all the couplings are taken to be constant
$V_{f}(k)=V_{b}(k)=V/\sqrt{2}$, because $\vert V_{f}(k)\vert^{2}$
and $\vert V_{b}(k)\vert^{2}$  { are much less than} $\omega_{0}$
and vary slowly around $k_{0}$. This   {is a direct result of} the
memoryless character of the photonic field, known as the Markov
approximation~\cite{RevModPhys.70.101,carmichaelQO}.

In the quantum-jump (or quantum-trajectory) approach, the
dissipative time evolution is described by a non-Hermitian
effective Hamiltonian~\cite{carmichaelQO,RevModPhys.70.101}:
\begin{equation}
  \hat H_{\rm eff }=\hat H-i\frac{\gamma}{2}\hat{\sigma}_{_{EE}},
 \label{Heff}
\end{equation}
where the imaginary term $i\gamma/2$, with the damping constant
$\gamma$, refers in our model to   a nondirectional spontaneous
emission of the excited state $\vert E\rangle$. The evolution,
governed by Hamiltonian~(\ref{Heff}), can be interrupted by random
quantum jumps. However, we study the conditional dynamics, which
is described only by the non-Hermitian Hamiltonian $\hat{H}_{\rm
eff }$ without quantum jumps,  {since the dissipation results in
detectable photon loss}~\cite{carmichaelQO}.

After transforming   the Hamiltonian~(\ref{Heff})  into the frame
rotating with input-photon frequency $\omega_{0}$, the integration
ranges in Eqs.~(\ref{Hamliltonian0}) and (\ref{Hamliltonian1}) can
be extended to $(-\infty,\infty)$ for performing the Fourier
transform of the field operators, because only a narrow bandwidth
in the vicinity of $\omega_{0}$ will be taken into consideration
and has a nontrivial contribution to the final scattering process.
The dynamics of the system   consisting of a $\Lambda$-type atom and
  a waveguide is described by the Hamiltonian $\tilde{H}_{\rm
eff}$ in real space as follows:
\begin{eqnarray}
\tilde{H}_{\rm eff}  &=&\tilde{H_{0}}+\tilde{H_{1}},\\
 \tilde{H}_{0} &=&\tilde{\triangle}\hat{\sigma}_{_{EE}}-i\int {\rm d}x[\hat{a}_{f}^{\dagger}(x)\partial_{x}\hat{a}_{f}(x)
 -\hat{a}_{b}^{\dagger}(x)\partial_{x}\hat{a}_{b}(x)],\nonumber \\
\tilde{H}_{1} &=&\!\!\int\!\!{}{\rm
d}x\delta(x)\!\!\left[\sqrt{\Gamma_{f}}\hat{a}_{f}^{\dagger}(x)\hat{\sigma}_{+{_E}}
+\sqrt{\Gamma_{b}}\hat{a}_{b}^{\dagger}(x)\hat{\sigma}_{-{_E}}+{\rm
h.c.}\right].\nonumber
\end{eqnarray}
Here $\delta(x)$ is the Dirac delta function modeling the
scattering point at $x=0$, where the atom is placed. The effective
detuning between the atom and an input photon is
$$\tilde{\triangle}=\omega_{_A}-\Omega_{_E}+i\gamma/2.$$ Moreover,
$\hat{a}_{f,b}(x)$ [$\hat{a}_{f,b}^{\dagger}(x)$] annihilates
(creates) a forward-propagating (backward-propagating) photon at the point
$x$. These operators are related to the corresponding operators in
the wave vector representation by the Fourier transforms:
\begin{eqnarray}
\hat{a}_{f}(x) & =&\frac{1}{2\pi}\int {\rm d}k\,\hat{a}_{f}(k)e^{ikx} \;,\nonumber \\
\hat{a}_{b}(x) & =&\frac{1}{2\pi}\int {\rm d}k\,\hat{a}_{b}(k)e^{-
ikx} \;.
\end{eqnarray}

To study the scattering of a single photon  {with} frequency
$\omega_{_A}$ by this atom-waveguide system, one should consider
two cases: (1) If a given input photon does \emph{not} match a
circularly-polarized transition of the atom, then it propagates in
the forward modes in the waveguide without any disturbance;
however, (2) if an input photon matches a given transition, then
it  {strongly} couples to the atom.  {Note that
forward-propagating (backward-propagating) modes evolve into the
circularly polarized state $\vert\sigma_{+}\rangle$
($\vert\sigma_{-}\rangle$) at the position of the atom.} Thus,
{if the atom is initially prepared in the state $\vert+\rangle$,
the quantum state of a forward-propagating photon changes
significantly due to scattering   from the atom.  Otherwise, it
passes through the atom without change.}

For calculating the reflection and transmission amplitudes, we
first give a general state of the single-excitation subspace of
the system composed of the atom and waveguide
modes~\cite{shen2005coherent}, which reads as
\begin{eqnarray}\label{ansanz}
\vert\psi_{1}\rangle &=&\phi_{_E}\vert\Phi,E\rangle\nonumber+\int
{\rm d} x \phi_{f}(x)\hat{a}_{f}^{\dagger}(x)\vert\Phi,+\rangle
\\
&&+\int {\rm
d}x\phi_{b}(x)\hat{a}_{b}^{\dagger}(x)\vert\Phi,-\rangle,
\end{eqnarray}
where $\hat{a}_{f}^{\dagger}(x)\vert\Phi\rangle$
[$\hat{a}_{b}^{\dagger}(x)\vert\Phi\rangle$] denotes a forward-propagating
{(backward-propagating)} photon at the point $x$ with
{single-photon wave function} $\phi_{f}(x)$ [$\phi_{f}(x)$], 
$\vert\Phi\rangle$ denotes the vacuum state for optical-field
operators, and $\phi_{_E}$ represents the probability amplitude
that the atom is excited to the state  $\vert E\rangle$. We can
then give a set of equations resulting from the time-independent
Schr\"{o}dinger equation $\tilde{H}_{\rm
eff}\vert\psi_{1}\rangle=\omega_{_A}\vert\psi_{1}\rangle$:
\begin{eqnarray}
0&=&\left(-i\partial_{x}-\omega_{_A}\right)\phi_{f}(x)+\sqrt{\Gamma_{f}}\delta(x)\phi_{_E}, \nonumber \\
0&=&\left(i\partial_{x}-\omega_{_A}\right)\phi_{b}(x)+\sqrt{\Gamma_{b}}\delta(x)\phi_{_E}, \nonumber \\
0&=&-\tilde{\triangle}\phi_{_E}(x)+\sqrt{\Gamma_{f}}\phi_{f}(0)+\sqrt{\Gamma_{b}}\phi_{b}(0).
\label{eq:schrodinger01} \label{eq:schrodinger1}
\end{eqnarray}
We assume the following ansatz for the solutions of the above
equations:
\begin{eqnarray}
&&\phi_{f}(x)=\exp{(ik_{\!f}x)}\left[\varTheta_H(-x)+T\varTheta_H(x)\right], \nonumber \\
&&\phi_{b}(x)=\exp{(-ik_{b}x)}R\varTheta_H(-x).
\end{eqnarray}
where $T$ and $R$ are the transmission and reflection probability
amplitudes of this atom-waveguide system, and $\varTheta_H(x)$ is
the Heaviside step function with $\varTheta_H(x)|_{x=0}=1/2$ and
$\frac{\partial{\varTheta_H(x)}}{\partial{x}}|_{x\rightarrow0^+}=1$.
The set of equations, given in Eq.~(\ref{eq:schrodinger01}),
evolves into
\begin{eqnarray}
0&=&\left(k_{f}\!-\!\omega_{_A}\right)\phi_{f}(x)\!-\!ie^{ik_{\!f}x}\delta(x)[T-\!1]\!+\!\sqrt{\Gamma_{f}}\delta(x)\phi_{_E},\nonumber\\
0&=&\left(k_{b}-\omega_{_A}\right)\phi_{b}(x)-ie^{-ik_{b}x}\delta(x)R+\sqrt{\Gamma_{b}}\delta(x)\phi_{_E},\nonumber\\
0&=&-\tilde{\triangle}\phi_{_E}+\sqrt{\Gamma_{f}}\phi_{f}(0)+\sqrt{\Gamma_{b}}\phi_{b}(0).
\end{eqnarray}
After inserting the solution ansatz into the above equations, 
we directly obtain
\begin{eqnarray}
\phi_{_E}&=&\frac{2\sqrt{\Gamma_{f}}}{2\widetilde{\Delta}+i(\sqrt{\Gamma_{b}\Gamma_{f}}+\Gamma_{b})},\nonumber \\
R & =& -i\sqrt{\Gamma_{b}}\phi_{_E},\nonumber\\
T & =& \frac{2\widetilde{\Delta}
+i(\Gamma_{b}-\sqrt{\Gamma_{b}\Gamma_{f}})}{2\sqrt{\Gamma_{f}}}\phi_{_E},\label{tcoefficient}
\end{eqnarray}
It could  easily be found that an input photon in a proper
direction is totally reflected with $R=-1$ and $T=0$ in the ideal
resonant scattering process, assuming no detuning,
$\Delta=\omega_{_A} -\Omega_{_E}=0$,  and the same coupling rates
$\Gamma_{b} =\Gamma_{f}=\Gamma$, which are much greater than the
damping rate, i.e., $\Gamma\gg\gamma$. This means that we assume
the strong-coupling regime~\cite{gu2017microwave}. At the same
time, the ground state of the atom is flipped into the other one.
That is, a perfect single-photon scattering process is completed
and the photon state and the ground states of the $\Lambda$-type
atom are flipped simultaneously.

\begin{figure}[!t]
\centering
\includegraphics[width=8.5 cm,angle=0]{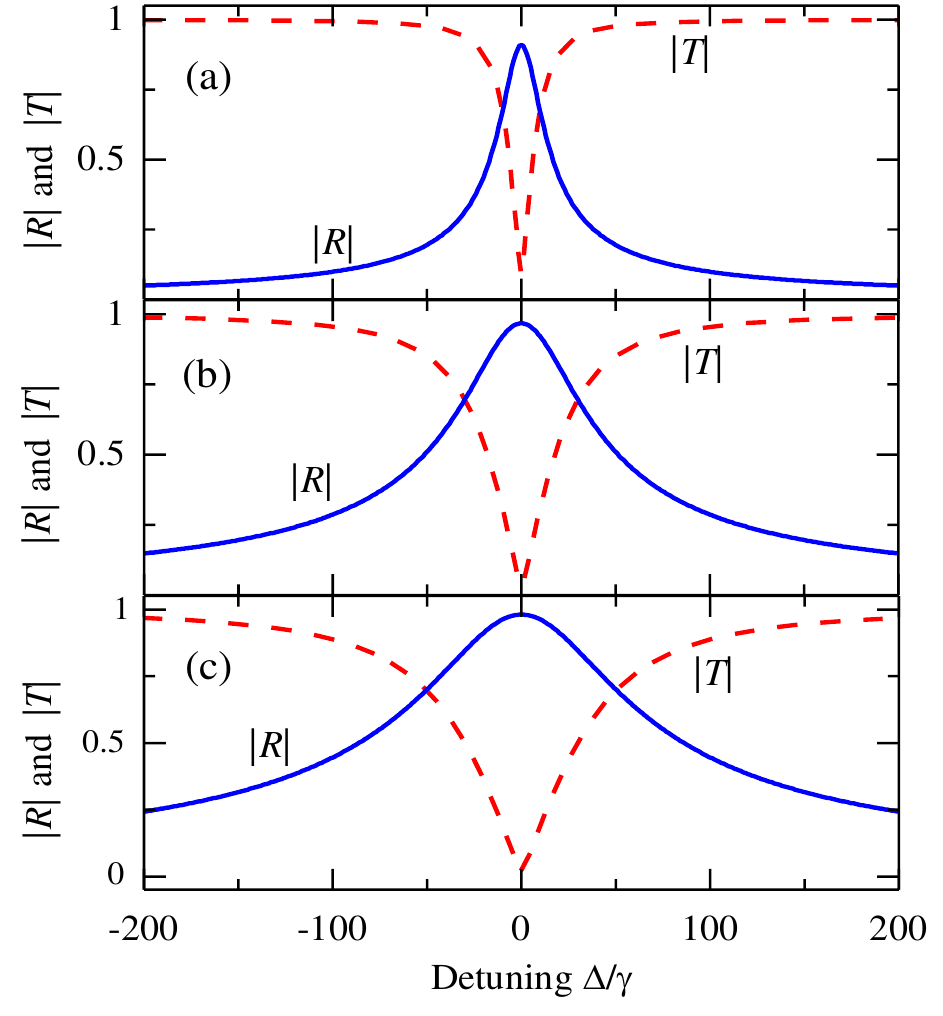}
\caption{Absolute values of the transmission   $|T|$ and reflection   $|R|$
amplitudes versus the detuning $\Delta$ (in units of the damping
rate $\gamma$) between an input photon and the atomic transitions
with (a) $\beta=10$, (b) $\beta=30$, and (c) $\beta=50$. Here
$\beta=\Gamma/\gamma$ is the ratio of the coupling constant
$\Gamma$ and the damping rate $\gamma$ describing the enhancement
of the directional emission of the atom.} \label{fig2}
\end{figure}

Before analyzing   a  possible application of the MQI for QIP, we first
study the transmission and reflection   amplitudes
versus   practical parameters: the photon detuning $\Delta$ and
the ratio $\beta=\Gamma/\gamma$ of the coupling and damping rates
showing the enhancement of the directional emission of the atom,
shown in Fig.~\ref{fig2}.
 {The absolute values of the transmission and reflection amplitudes, $|T|$ and $|R|$,
in Eq.~(\ref{tcoefficient}) are obtained from} the stationary solutions
in   real space.
Furthermore, we numerically verify these
  amplitudes $T$ and $R$ by solving the time-dependent
Schr\"{o}dinger equation $-i\frac{\partial\vert\psi\rangle
}{\partial t}=\hat{H}_{\rm eff}\vert\psi\rangle$, where
$\vert\psi\rangle$ denotes a general single-excitation state  in
the wave-vector space and evolves under the wave-vector space
Hamiltonian $\hat{H}_{\rm eff}$ in Eq.~(\ref{Heff}), and the
wave-vector distribution of an input photon is assumed to be a
Gaussian pulse,
\begin{eqnarray}
f(k)=\frac{5}{k_0\sqrt{\pi}}\exp[-(k-k_0)^2 (5/k_0)^2].
\end{eqnarray}
On the scale of Fig.~\ref{fig2}, there is no difference between
the analytical results and the corresponding numerical
calculations.

It is clearly seen that there are three distinct detuning points
for   a single-photon scattering   {process,} involving the
atom-waveguide discussed here. One of those is the resonant point,
where the input single photon is almost deterministically
reflected with the atomic ground state flipped, and the largest
difference is observed between the reflection and transmission
modes. While at the other two points for the detuning
$\Delta\thickapprox\pm\Gamma$, the scattered photon propagates
with equal probability in the reflection and transmission modes
and the atom is projected into the corresponding ground state,
which leads to   an entanglement between the ground states of the
$\Lambda$-type atom and the photon state~\cite{togan2010quantum}.

\section{Multifunctional quantum interface for QIP}\label{sec3}

The proposed system,   consisting of a $\Lambda$-type atom and a
nanowaveguide, provides   a deterministic   quantum interface between
a single photon and a single atom. When a photon is injected in
the direction that excites a  local circularly polarized field
that the atom is coupled   to, then the photon is either
reflected along with a ground-state flip of the atom, or keeps
propagating in the transmission modes,   leaving the atom without
changing its original state. The relative probabilities of these
two scattering modes can be controlled by the detuning $\Delta$.
In this section, we describe   a MQI as an elementary building
block of   chiral quantum networks. The SWAP gate, the quantum memory,
and a hybrid-entangling gate  between a single atom and a
polarized-encoded single photon could, in principle, be
implemented in the same setup that is trigged only when a
polarized-encoded photon is input in a particular port. When a
photon enters the MQI from the other port, it passes through the
setup with a polarization flip, leaving  the atom unchanged. This
provides  potential applications for  chiral quantum networks when
multiple channels, linking different nodes, are
tunable~\cite{PhysRevA.91.042116}.

\subsection{General scattering output matrix}

\begin{figure}[!t]
\includegraphics[width=0.95\linewidth]{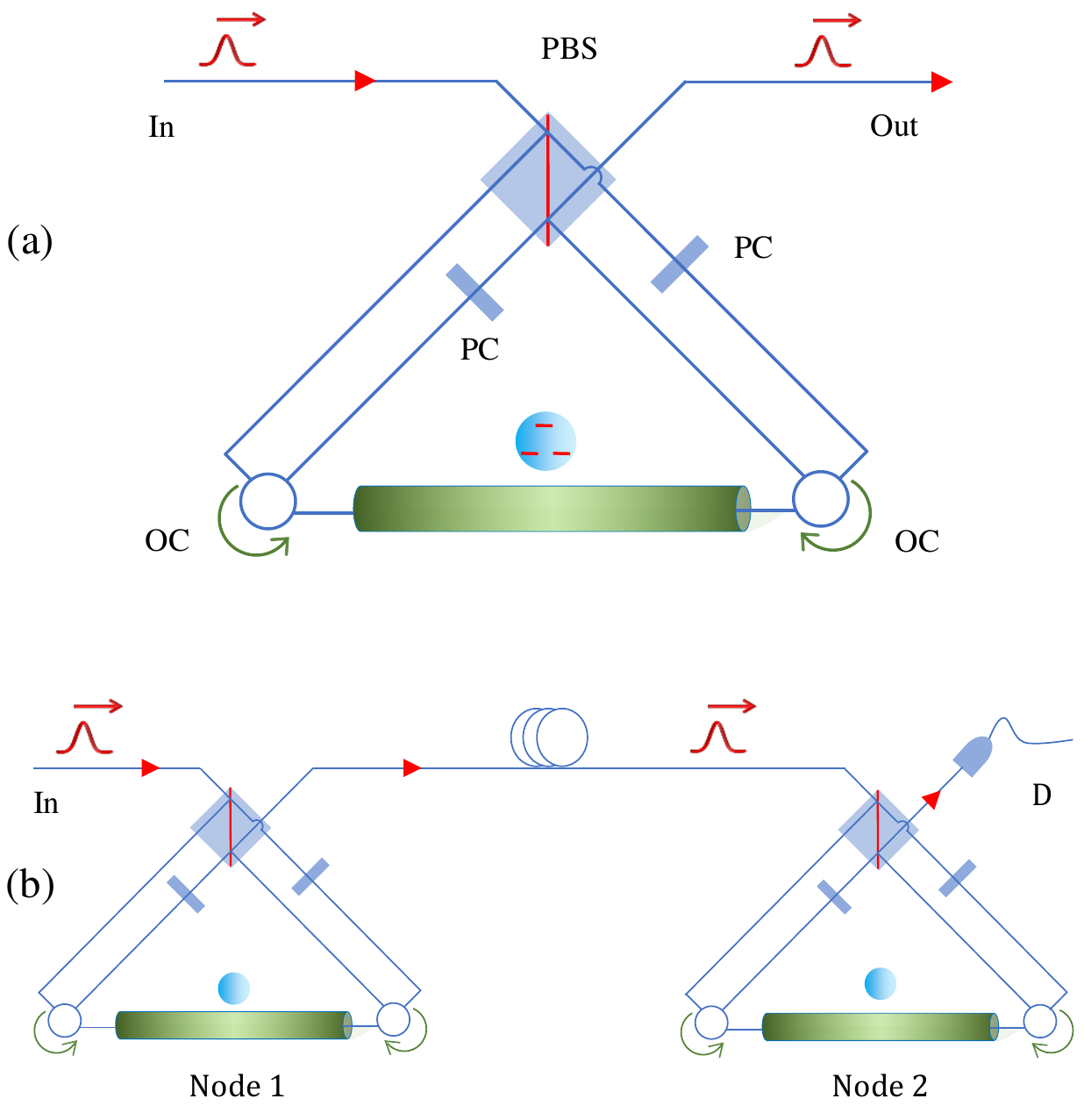}
\caption{(a) Schematics of the multifunctional quantum interface
(MQI)  { using a $\Lambda$-type emitter and a 1D }
waveguide and (b) heralded-entanglement generation and quantum-state transmission between two remote quantum nodes. Here,   PBS
denotes a polarized beam splitter, OC is an optical circular, and
PC is a polarization controller.} \label{fig3}
\end{figure}

The setup for chiral QIP with polarized-encoded single photons is
shown in Fig.~\ref{fig3}(a). It is composed of a dominant
nanowaveguide coupled to a $\Lambda$-type atom and several linear
optical elements: a polarized beam splitter (PBS),  an optical
circulator (OC), and a polarization controller (PC). The PBS
transmits  horizontally polarized photons (state $\vert H\rangle$)
and reflects  vertically polarized photons (state $\vert
V\rangle$).
 {The PC flips the polarization of a photon passing through it and performs
the conversion $\vert H\rangle \leftrightarrow |V\rangle $, while the OC is used to spatially  separate the input and output modes~\cite{duan2004scalable, koshino2010, reiserer2014quantum} by
transmitting an input photon in one port to an output port determined by the  operation direction of the circulator}.

{A single photon encoded in the  superposition polarization state}
$\vert\psi_{_P}\rangle=\alpha_{_H}|H\rangle+\alpha_{_V}|V\rangle$
is split into two directions  by the PBS, followed by a
polarization flip on the transmission modes. Therefore, a single
photon passing through either the right or left OC has the same
polarization, which maximally couples the photon into   a
waveguide from either direction and creates a right-
($\vert\sigma_{+}\rangle$) or left- ($\vert\sigma_{-}\rangle$)
circularly polarized local field at the atom
position~\cite{le2015nanophotonic,guimond2016chiral,PhysRevA.91.042116,PhysRevA.83.021803,bliokh2014extraordinary, bliokh2015quantum,
bliokh2015spin, bliokh2015transverse,lodahl2017chiral}. For the atom
initialized in the state
$\vert\psi_{_A}\rangle=\beta_{+}|+\rangle+\beta_{-}|-\rangle$, the
combined atom-photon state evolves into
\begin{eqnarray}
 \vert\psi_1\rangle=&&\beta_{+}\left(T\alpha_{_V}\vert\sigma_{+},+\rangle+R\alpha_{_V}\vert\sigma_{-},-\rangle\right)\nonumber\\
 && +\;\beta_{-}\left(T\alpha_{_H}\vert\sigma_{-},-\rangle+R\alpha_{_H}\vert\sigma_{+},+\rangle\right)\nonumber\\
 && +\;\beta_{+}\alpha_{_H}\vert\sigma_{-},+\rangle+\beta_{-}\alpha_{_V}\vert\sigma_{+},-\rangle.
\end{eqnarray}
According to the time-reversal symmetry of the waveguide, the
polarization   of a photon  evolves into the vertically polarized
state $|V\rangle$, and it is rerouted by the OCs followed
by a polarization flip on the left-propagating modes.
Subsequently, the two propagating modes of the photon recombine
together at the PBS, and the combined system evolves into the
state $|\psi_2\rangle$, when the photon is leaving the PBS:
\begin{eqnarray}
 \vert\psi_2\rangle
  =&&\beta_{+}\vert H\rangle\left(R\alpha_{_V}\vert-\rangle+\alpha_{_H}\vert+\rangle\right)\nonumber\\
 &&+\;\beta_{-}\vert V\rangle\left(\alpha_{_V}\vert-\rangle+ R\alpha_{_H}\vert+\rangle\right)\nonumber\\
  &&+\;T\left(\beta_{+}\alpha_{_V}\vert V,+\rangle+\beta_{-}\alpha_{_H}\vert H,-\rangle\right).
\end{eqnarray}
This process could  be described by a general scattering matrix
$\hat{S}$ that connects the initial and final states of the
combined system consisting of a polarized photon and a
$\Lambda$-type atom as
\begin{eqnarray}
 \vert\psi_2\rangle=\hat{S} \vert\psi_{_A}\rangle \vert\psi_{_P}\rangle,
 \label{outputp}
\end{eqnarray}
where
\begin{eqnarray}
\hat{S}=
\begin{pmatrix}
 1 & 0 & 0 & 0\\
0 & T & R & 0\\
0 & R & T & 0\\
0 & 0 & 0 & 1\\
\end{pmatrix}.\label{smatrix1}
\end{eqnarray}

\subsection{Quantum gates and memory}

In general, specific functions of the MQI depend on the
{reflection and transmission} amplitudes, $R$ and $T$, which are
tunable by controlling the detuning between an input photon and
the atom.  {By properly selecting scattering conditions, the
atom-waveguide system evolves as intended, leading to desired
functions of the MQI.}

\subsubsection{SWAP gate}

As discussed in   the previous section, when the frequency of an input
photon   is nearly  {resonant with} the atom,   the photon is
deterministically reflected by the atom.  {Simultaneously,}  the state of the atom is flipped, with $R=-1$ and $T=0$ for an ideal
directional scattering process,    if the coupling constant is much
greater than the damping rate, i.e., $\Gamma\gg\gamma$, which
corresponds to the strong-coupling regime. Therefore, the general
scattering matrix $\hat{S}$ is simplified to
\begin{eqnarray}
\hat{S}_{\Delta=0}=
\begin{pmatrix}
 1 & 0 & 0 & 0\\
0 & 0 & -1 & 0\\
0 & -1 & 0 & 0\\
0 & 0 & 0 & 1\\
\end{pmatrix}.\label{smatrix2}
\end{eqnarray}
 {Obviously, this scattering process equivalently performs}
the standard SWAP gate between a polarized
photon and a single atom, because the $\pi$-phase shift could be
completely compensated by local phase-flip operations
$\hat\sigma_z$:
\begin{eqnarray}
  {\rm SWAP} &=& (\hat I\otimes\hat\sigma_z) \hat{S}_{\Delta=0} (\hat
  I\otimes\hat\sigma_z),
\label{SWAP}
\end{eqnarray}
where $\hat I$ is the qubit identity operator.

\subsubsection{Quantum memory}

In order to perform the quantum memory  {of a single photon
encoded in an arbitrary polarization state,} one can first perform
the SWAP gate, as described above, and then perform a measurement
of the scattered photon to guarantee that the SWAP gate has been
completed and the state of the polarized photon is now stored in
the atom. After a while, to read out the state of the original
photon, one can perform another SWAP gate by impinging another
photon. Although the second photon could, in principle, be
initially in an arbitrary pure state or even a mixed state;   here,
for simplicity, we assume that the initial polarized photon state
is $|V\rangle$. The atom, in an ideal scattering process, is
mapped into the ground state $|-\rangle$ and a photon of the same
polarization, as that of the original photon, is generated
simultaneously.  {Thus, these operations store and retrieve a
single-photon state, which completes the quantum memory
procedure.}

\subsubsection{$\sqrt{{\rm SWAP}}$ gate}

The setup shown in   Fig.~\ref{fig3}(a) could also be exploited
to generate    entanglement between a polarized photon and a
single atom, which can be referred to as hybrid entanglement.
Instead of working near the resonant point, one   introduces a
detuning of ($\pm\Gamma$) for a given input photon. Now the
general scattering matrix, shown in Eq.~(\ref{smatrix1}),  is
specified as follows:
\begin{eqnarray}
\hat{S}_{\Delta=\Gamma}=
\begin{pmatrix}
 1 & 0 & 0 & 0\\
0 & -\frac{i-1}{2}& -\frac{1+i}{2} & 0\\
0 & -\frac{1+i}{2} & -\frac{i-1}{2} & 0\\
0 & 0 & 0 & 1\\
\end{pmatrix}\label{smatrix3}
\end{eqnarray}
and
\begin{eqnarray}
\hat{S}_{\Delta=-\Gamma}=\hat{S}^{-1}_{\Delta=\Gamma}=
\begin{pmatrix}
 1 & 0 & 0 & 0\\
0 & \frac{1+i}{2}& -\frac{1-i}{2} & 0\\
0 & -\frac{1-i}{2} & \frac{1+i}{2} & 0\\
0 & 0 & 0 & 1\\
\end{pmatrix}.\label{smatrix4}
\end{eqnarray}
By applying $\hat{S}_{\Delta=\Gamma}$ or
$\hat{S}_{\Delta=-\Gamma}$ on a hybrid system consisting of a
single atom and a polarized photon,
$|\psi_A\rangle|\psi_P\rangle$, the square root of the SWAP gate
($\sqrt{\textrm{SWAP}}$) is   accomplished in this system, because
\begin{equation}
  (\hat{S}_{\Delta=\Gamma})^2= (\hat{S}_{\Delta=-\Gamma})^2 =
  \hat{S}_{\Delta=0}.
 \label{SWAP2}
\end{equation}
We note that our gate, given in Eq.~(\ref{smatrix4}), differs from
the standard $\sqrt{\textrm{SWAP}}$ gate only by local
operations, i.e.,
\begin{eqnarray}
  \sqrt{\textrm{SWAP}} &=& (\hat I\otimes\hat\sigma_z)
  \hat{S}_{\Delta=-\Gamma} (\hat I\otimes\hat\sigma_z).
\label{SqrtSWAP}
\end{eqnarray}
Thus, our system can be used to generate the maximal entanglement
between a polarized photon and a single atom, by   initializing these
to be in the state $|H\rangle|+\rangle$ or $|V\rangle|-\rangle$.
When combined with the classical swapping procedure described
above, one can establish the maximal entanglement between two
remote atoms placed at different nodes of quantum networks, as
shown in Fig.~\ref{fig3}(b).

So far, we have described the SWAP gate, the quantum memory, and  {
the hybrid-entangling $\sqrt{\textrm{SWAP}}$  gate,} based on the
effective scattering of the atom-waveguide system. However, if one
switches the output port of the MQI to be a new input port, one
can easily find that the probability amplitude of the input
polarized photon is first divided into two amplitudes of
propagating modes, and then the photon exhibits a polarization
flip. These modes are directly transmitted to the PBS by any of
the OCs. Finally, the photon reaches the PBS with a polarization
flip from the original input port of the MQI. In summary, when a
photon enters at a particular input port, it enables   the MQI to
perform various QIP tasks, while the photon enters from the other
input port then just passes through the MQI leaving the atom
unaffected. This provides an interesting application for   complex
quantum networks when multiple connecting channels are available,
because the quantum function of the MQI could be switched on or
off for various QIP tasks simply by choosing different connecting
channels.

\section{Fidelity and efficiency of QIP with MQI}\label{sec4}

In   the  previous section, we described the MQI for  unidirectional
implementations of the SWAP and $\sqrt{\textrm{SWAP}}$ gates, and
quantum memory, for all   the tasks which, in principle,   work in
a deterministic way. However, the practical scattering output of
the MQI can deviate from the ideal outcome due to  the finite
linewidth of   the input photon and   the nondirectional spontaneous
emission of the atom. Here we use both   the average fidelity $\bar{F}$
and the average efficiency $\bar{\eta}$ to evaluate the performance of
the MQI for different QIP tasks.

For the SWAP and quantum hybrid-entangling $\sqrt{{\rm SWAP}}$
gates, the fidelities and efficiencies depend on the states of the
atom and the photon involved. To evaluate the performance of these
processes, one can use the average fidelity and
efficiency~\cite{ritter2012elementary,
nielsen2002simple,rong2015experimental}. The average fidelity
$\bar{F}$ is defined as the average overlap between the ideal and
the practical  outputs for different photonic and atomic states.
Thus, the average fidelities for the SWAP and $\sqrt{{\rm SWAP}}$
gates read, respectively, as
\begin{eqnarray}
\bar{F}_{\rm swap}
&=&\frac{1}{N}\sum_{i=1}^{N}\sqrt{\frac{\vert\langle\psi_i|\hat{S}^{\dagger}_{\Delta=0}|\psi_{\rm
swap}^{(i)}
 \rangle\vert^2}
{\langle\psi_{\rm swap}^{(i)}|\psi_{\rm swap}^{(i)}\rangle}},\\
\bar{F}_{\rm ent} &=&
\frac{1}{N}\sum_{i=1}^{N}\sqrt{\frac{\vert\langle\psi_i|\hat{S}^{\dagger}_{\Delta=\Gamma}|\psi_{\rm
ent}^{(i)}\rangle\vert^2} {\langle\psi_{\rm ent}^{(i)}|\psi_{\rm
ent}^{(i)} \rangle}},
\end{eqnarray}
where $\hat{S}_{\Delta=0}$ and $\hat{S}_{\Delta=\Gamma}$ are the
ideal transition matrices given, respectively, in
Eqs.~(\ref{smatrix2}) and~ (\ref{smatrix3}), and $|\psi_{\rm
swap}^{(i)}\rangle$ ($|\psi_{\rm ent}^{(i)}\rangle$) is the
corresponding output state obtained from the frequency-dependent
scattering matrix resulting from Eq.~(\ref{smatrix1}). Moreover,
$|\psi_i\rangle$ is a state chosen from the overcomplete set of
$N=36$ states:
 $\{|+\rangle$,$|-\rangle$, $(|+\rangle\pm|-\rangle)/\sqrt{2}$, $(|+\rangle\pm i|-\rangle)/\sqrt{2}\}$
 $\otimes \{|H\rangle$, $|V\rangle$, $(|H\rangle\pm|V\rangle)/\sqrt{2}$, $(|H\rangle\pm i|V\rangle)/\sqrt{2}\}$,
where
\begin{equation}
  |j\rangle=\int{}d\omega{}f(\omega)\hat{a}_{j}^{\dagger}(\omega)|\Phi\rangle,
 \label{j-state}
\end{equation}
where ($j=H,V$) represents a single-photon pulse of the frequency
distribution $f(\omega)$.

When the MQI is used to perform   the quantum memory for an arbitrary
single-photon pulse, an initially separable state of the combined
system,   consisting of a single photon and an atom, is   modified due to
  its  nonzero transmission in a practical scattering process. In
order to increase the fidelity of the quantum memory, one can  {initially prepare the atom }
in a deterministic state $|-\rangle$. After
the storage and read-out processes, the measurement on the atom in
the basis $\{|+\rangle,|-\rangle\}$ is performed. The average
fidelity conditioned on the measurement of the atom to the state
$|-\rangle$ is given by
\begin{eqnarray}
\bar{F}_{\rm mem}
=\frac{1}{N}\sum_{i=1}^{N}\sqrt{\frac{\vert\langle\phi_i|\psi_{\rm
mem}^{(i)}\rangle\vert^2} {\langle\psi_{\rm mem}^{(i)} |\psi_{\rm
mem}^{(i)}\rangle}},
\end{eqnarray}
where $N=6$; $|\phi_i\rangle$ is chosen from the overcomplete
set of states $\{|H\rangle$, $|V\rangle$,
$(|H\rangle\pm|V\rangle)/\sqrt{2}$, $(|H\rangle\pm
i|V\rangle)/\sqrt{2}\}$; and $|\psi_{\rm
mem}^{(i)}\rangle=\hat{M}|\phi_i\rangle$ is the realistic output
state with the following frequency-dependent filter matrix: $\hat{M}=\binom{R^2 ~0}{ T ~1}$.

\begin{figure}[tbp]
\includegraphics[width=8.5cm]{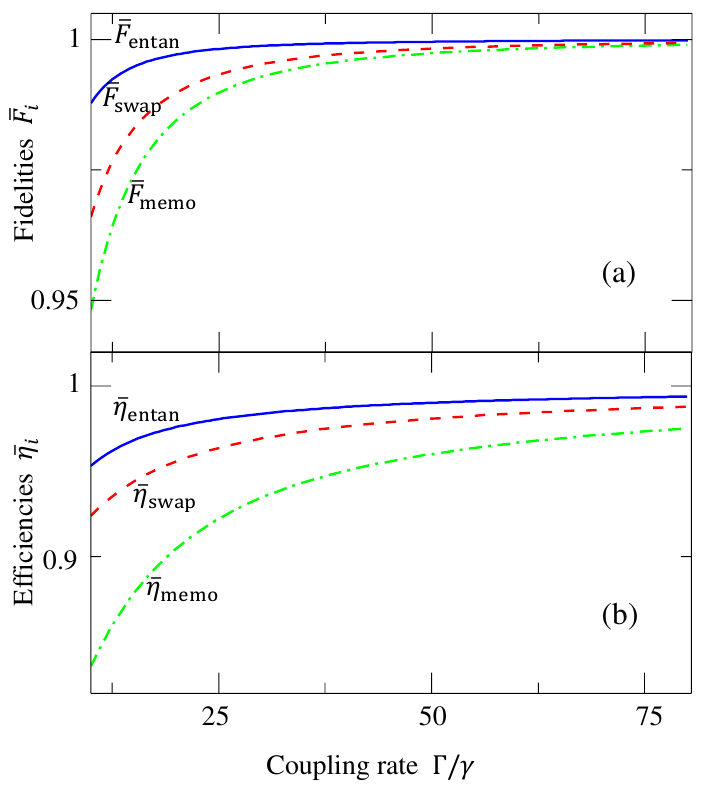}
\caption{(a) The average fidelities $\bar{F}_{\rm ent}$,
$\bar{F}_{\rm swap}$, and $\bar{F}_{\rm mem}$ and (b) the average
efficiencies $\bar{\eta}_{\rm ent}$,  $\bar{\eta}_{\rm swap}$, and
$\bar{\eta}_{\rm mem}$ versus the coupling constant $\Gamma$ (in
units of the damping rate $\gamma$). The averages are calculated
over all detunings of an input photon, with the Gaussian wave form
given by Eq.~(\ref{gaussianPuls})  with $\sigma_{\omega}=5\gamma$.
} \label{figsigma}
\end{figure}

\begin{figure}[tbp]
\includegraphics[width=8.5cm]{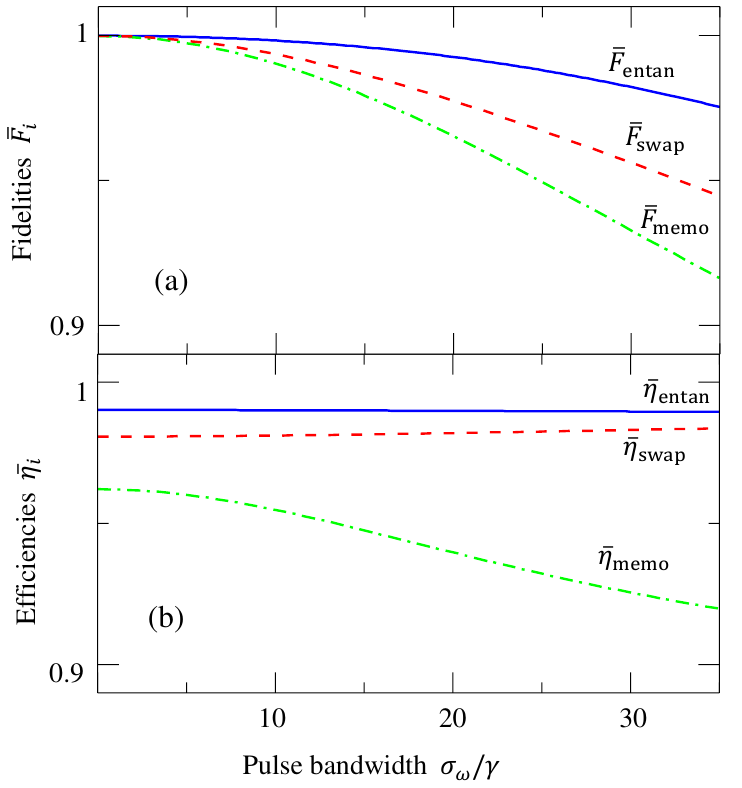}
\caption{(a) Average fidelities $\bar{F}_{\rm ent}$, $\bar{F}_{\rm
swap}$, and $\bar{F}_{\rm mem}$ and (b) average efficiencies
$\bar{\eta}_{\rm ent}$, $\bar{\eta}_{\rm swap}$, and
$\bar{\eta}_{\rm mem}$ versus the pulse bandwidth
$\sigma_{\omega}/\gamma$ (in units of the damping rate $\gamma$).
These averages are calculated over all detunings of an input
photon, with the Gaussian wave form given by
Eq.~(\ref{gaussianPuls})  and $\Gamma/\gamma=50$. }
\label{figGamma}
\end{figure}

Moreover, the average efficiencies $\bar{\eta}_k$ can be given by
the average probabilities of observing directional scattering,
i.e.,
\begin{eqnarray}
 \bar{\eta}_{k}&=&\frac{1}{N}\sum_{i=1}^{N}{\langle\psi_{k}^{(i)}|\psi_{k}^{(i)}\rangle}.
\end{eqnarray}
In special cases, this formula reduces to $\bar{\eta}_{\rm swap}$,
$\bar{\eta}_{\rm ent}$, and $\bar{\eta}_{\rm mem}$, which are
shown in Figs.~\ref{figsigma} and~\ref{figGamma}.


The performance of the quantum gates and memory via the MQI are
shown in Fig.~\ref{figsigma}, where we plotted the fidelities and
the efficiencies  versus the effective directional scattering rate
$\Gamma/\gamma$ for a given Gaussian single-photon pulse  defined   by the spectrum
\begin{eqnarray}
f(\omega)=\frac{1}{\sqrt{\pi}\sigma_{\omega}}\exp\left[-\left(\frac{\omega-\omega_c}{\sigma_{\omega}}\right)^2\right],\label{gaussianPuls}
\end{eqnarray}
where $\omega_c$ is the center frequency and $\sigma_{\omega}$
denotes the pulse bandwidth. In general, the MQI implements the
SWAP and $\sqrt{\textrm{SWAP}}$ gates better than   the quantum memory.
This is because   only one non-ideal single-photon scattering
process is involved in our implementation of the gates, while two
scattering processes are involved for realizing   a quantum memory.
For the $\sqrt{\textrm{SWAP}}$ gate, both   the average fidelity and
average efficiency approach their respective stable values of
$\bar{F}_{\rm
ent}=0.9996$ and $\bar{\eta}_{\rm
ent}=0.9901$, when the directional
scattering rate is $\Gamma/\gamma \geq 50$ for a narrow
input-photon pulse $\sigma_{\omega}=5\gamma$. For   the SWAP gate, the
average fidelity and efficiency are almost as high as those for
the $\sqrt{\textrm{SWAP}}$ gate,   with
 $\bar{F}_{\rm
swap}=0.9982$ and $\bar{\eta}_{\rm swap}=0.9810$,   for
$\Gamma/\gamma \geq 50$. {The fidelity and efficiency of the
quantum memory are slightly lower} than those for the two gates.
  The average fidelity and average efficiency of   the
quantum memory are still large enough for a practical quantum
network, because $\bar{F}_{\rm mem} \geq 0.9928$ and
$\bar{\eta}_{\rm mem}\geq 0.9345$ are achievable for the effective
directional scattering rate $\Gamma/\gamma \geq 30$.

To study the influence of the bandwidth of an input photon, the
performances of all three functions  of the MQI are shown in
Fig.~\ref{figGamma} for the ratio of the directional coupling
constant  and damping rate, $\Gamma/\gamma=50$. In general, the
average fidelities of all three functions decrease
with increasing the bandwidth $\sigma_{\omega}/\gamma$ of the
single-photon pulse. This is because  the increment of
$\Gamma/\gamma$   leads to a larger deviation from the ideal
scattering condition for each quantum process. However, this
larger deviation  {compensates partially other detrimental effects, and
contributes to} the average efficiency
of the SWAP and $\sqrt{\textrm{SWAP}}$ gates, because the wide
bandwidth leads to a larger detuning at the edge range that
contributes less to   a realistic scattering, resulting in a smaller
absorbtion for the subsequent nondirectional emission.
Furthermore, the MQI performs excellently for the three QIP tasks
  when $\sigma_{\omega}/ \gamma\leq 25$.  The lowest performance among
these three tasks is for  the quantum memory, which has   an
average fidelity of $\bar{F}_{\rm mem}\geq 0.95$ and   an average
efficiency of $\bar{\eta}_{\rm mem}\geq 0.92$.

 {So far we have assumed that the optical circulator (OC) has
perfect efficiency. Nevertheless, realistic OCs always cause some
small losses of photons passing through
them~\cite{HallPRL2011Switching}. For example, a waveguide-based
OC with a loss rate of 0.05~dB has been designed and
simulated~\cite{takei2010circulator}. Moreover,
photon losses introduced by OCs only slightly reduce the fidelity
of an MQI, because an OC reduces equally both polarization
components of a single photon. In this respect, an OC can be
replaced by an unbalanced beam splitter~\cite{hacker2016photon, sun2016quantum}.}

 {To date, single-photon sources of narrow bandwidth based on
both natural and artificial atoms have been significantly
improved~\cite{kuhn2002deterministic, keller2004continuous,
holleczek2016quantum,aharonovich2011diamond}.  Nevertheless,
solid-state single-photon sources still suffer from spectral
diffusion and dephasing caused by phonons. These effects limit the
application of such photons for practical QIP
tasks~\cite{aharonovich2011diamond}. Fortunately, such spectral
broadening can be partially suppressed at low temperatures and
by coupling emitters to optical cavities or waveguides. For
example, a Fourier-transform-limited source based on NV centers in
nanocrystals has been demonstrated with a
linewidth of 16~MHz at cryogenic
temperatures~\cite{ShenzorophonoPRB2008}. Narrowband single photons could also be efficiently generated by a three-level emitter, strongly coupled to a high-finesse optical cavity, with vacuum-stimulated Raman transitions~\cite{kuhn2002deterministic, keller2004continuous,
holleczek2016quantum}. Furthermore, a solid-state single-photon source of subnatural linewidth has been demonstrated by operating in the small Rabi frequency limit of resonance fluorescence~\cite{Matthiesen2012narrow}.}

The dominant element of the MQI is the chiral  interaction between
a $\Lambda$-type atom and a single photon in a waveguide. A
promising implementation is a single negatively charged NV center coupled to a
photonic-crystal waveguide.
 {Here the states $\vert \pm\rangle$ of the $\Lambda$-level
structure, shown in Fig.~1(b), are the ground states of the NV
center with magnetic numbers $m_s=\pm 1$ associated with the orbital
angular momentum $m_l=0$.  The excited state $|E\rangle$ is given
by the state
$$|A_2\rangle=\frac{1}{\sqrt{2}}(|+1\rangle|E_-\rangle+|-1\rangle|E_+\rangle)$$
of the NV center, where $|E_{\pm}\rangle$ are the orbital states
with angular momentum $m_l=\pm1$, and $|{\pm 1}\rangle$ are the
spin states with magnetic numbers
$m_s=\pm1$~\cite{togan2010quantum,PhysRevB.92.081301,
PhysRevLett.114.053603,PhysRevLett.117.015502}.}
According to the total angular momentum conservation, the
transition $\vert +\rangle\leftrightarrow|E\rangle$ ($\vert
-\rangle\leftrightarrow|E\rangle$) is accompanied by the
absorption or radiation of a single $\sigma_+$ ($\sigma_-$)
polarized photon, and it has been used to generate quantum
entanglement involving polarized
photons~\cite{togan2010quantum,PhysRevB.92.081301}.

A photonic-crystal waveguide is a well-developed one-dimensional
system for enhancing the effective interaction between a flying
photon and a solid-state emitter~\cite{lodahl2015interfacing}. A
strong {coupling} between an emitter and a waveguide with
$\Gamma/\gamma\simeq50$ has been demonstrated~\cite{Arcari14}. A
waveguide photon can also be collected by a grating, a
tapered-mode adapter, or even a fiber for subsequent
operations~\cite{Arcari14,
tiecke2015efficient,PhysRevApplied.8.024026,PhysRevLett.117.240501}.
By using glide-plane waveguides, the maximum interaction between
an atom and a waveguide is achieved for unidirectional scattering,
because a chiral point corresponds to the field maximum of the
waveguide~\cite{PhysRevLett.117.240501, Mahmoodian:17}.   Moreover,
regular nanobeam waveguides and nanofibers coupled to NV center or
other three-level $\Lambda$-type emitters could also be used to
build this MQI~\cite{PhysRevLett.118.223603,
PhysRevX.7.031040,riedrich2014}.

\section{Discussion and summary}\label{sec5}

The chiral interaction between a single photon in a waveguide  and
a quantum emitter has   gained   considerable  attention
\cite{lodahl2017chiral,guimond2016chiral,PhysRevA.91.042116,young2015polarization,sollner2015deterministic}.
It combines the advantages of both   a chiral waveguide and   a
controllable $\Lambda$-type emitter system. The
spin-momentum-locked light in the chiral waveguide changes its
polarization when its  direction is
reversed~\cite{bliokh2014extraordinary,bliokh2015quantum,
bliokh2015spin,bliokh2015transverse}. This leads to a totally
different scattering when it is coupled to single emitters with
polarization-dependent dipole transitions. That is, either {zero}-
or $\pi$-phase shift occurs on the transmitted modes conditioned
on whether a propagating photon is coupled to the dipole
transition. Recently, deterministic entangling gates on single
photons and single emitters~\cite{young2015polarization,
sollner2015deterministic,PhysRevLett.117.240501} have been
proposed with negatively charged quantum dots which have two
independent dipole transitions and the initialization of this
system usually takes   a much longer time. For a $\Lambda$-type
emitter, it could, in principle, be initialized to an arbitrary
state determined by the state of an input photon, because a
passive quantum swapping exchanges the states of an input photon
and the emitter. Moreover, arbitrary single-qubit rotations on
both a stationary emitter and a polarized input photon are easily
achievable by faithful  optical control and passive linear-optical
elements. Furthermore, when the $\Lambda$-type atom is initially
in the excited state $|E\rangle$, the circular-dipole transitions
generate a maximally entangled state between the atom and a photon
encoded in the propagating modes~\cite{le2015nanophotonic}.

Our MQI combines the advantages of   chiral quantum optics and $\Lambda$-type
systems, and it has the following two important merits. (i) If an input photon is \emph{resonantly} coupled to
the dipole transitions, then the destructive interference of the
incident and transmitted modes of the photon leads to   a complete
photon reflection accompanied by a ground-state flip of the
atom, analogously to the one observed for the Sagnac
interferometers~\cite{bradford2012single}.
However, (ii) if an input photon is \emph{nonresonantly} coupled
to the transitions, then it can be both reflected and transmitted
with some nonzero probabilities. In particular, if the detuning
$\Delta$ is equal to ($\pm\Gamma$), then the reflection and
transmission probabilities are equal to each other. This generates
a maximally-entangled state between the atom and the photon. This
entanglement can be transferred (via transmitted photons and using
the SWAP gates) into the entanglement between two different
quantum nodes of quantum repeater-based
networks~\cite{dur1999quantum,
jiang2009quantum,sheng2013hybrid,munro2012quantum}.

In summary, we have presented an effective atom-photon interface
with a chiral waveguide. The tunable scattering process is
exploited to design the MQI for quantum networks, such as quantum
swapping and hybrid-entanglement generation between a single atom
and a single-photon pulse, which then leads to direct layouts of
quantum memory and the entanglement between different nodes. A
high performance of   this proposed MQI is within reach   of the existing
experimental   capabilities. Moreover, both the fidelity and the efficiency of
the MQI are robust to the potential imperfections originating from
a finite bandwidth of a single photon and a finite effective
coupling constant between the atom and a waveguide, which makes
our proposal useful for realistic quantum networks.

\section{Acknowledgments}

This work was supported in part by the: MURI Center for Dynamic Magneto-Optics via the Air Force Office of Scientific Research (AFOSR) (FA9550-14-1-0040), Army Research Office (ARO) (Grant No. 73315PH), Asian Office of Aerospace Research and Development (AOARD) (Grant No. FA2386-18-1-4045), Japan Science and Technology Agency (JST) (the ImPACT program and CREST Grant No. JPMJCR1676), Japan Society for the Promotion of Science (JSPS) (JSPS-RFBR Grant No. 17-52-50023), RIKEN-AIST Challenge Research Fund, and the John Templeton Foundation. 
T. L. thanks Yuran Zhang for helpful discussions. 
X.H. also acknowledges support by US ARO via Grant No. W911NF1710257.
K.X. acknowledges the support of the National Key R\&D Program of China
 (Grant No. 2017YFA0303703) and the ``1000 Young Talent" program.

\section*{References}
%





%

\end{document}